\documentclass[twocolumn]{aastex63}
\usepackage{graphicx}	
\usepackage{amsmath}	
\usepackage{amssymb}	
\usepackage{color}
\usepackage{array}
\usepackage{lineno}
\usepackage{bookmark}
\usepackage{subfigure}
\usepackage{booktabs}
\usepackage{float}
\usepackage{threeparttable}

\begin{document}

\title{Depolarization by jet precession in early optical afterglows of gamma-ray bursts}

\correspondingauthor{Tong Liu}
\email{tongliu@xmu.edu.cn}

\author[0000-0002-4448-0849]{Bao-Quan Huang}
\affiliation{Department of Astronomy, Xiamen University, Xiamen, Fujian 361005, China}

\author[0000-0001-8678-6291]{Tong Liu}
\affiliation{Department of Astronomy, Xiamen University, Xiamen, Fujian 361005, China}

\author{Guo-Yu Li}
\affiliation{Laboratory for Relativistic Astrophysics, Department of Physics, Guangxi University, Nanning 530004, China}

\begin{abstract}
Polarization observations provide a unique way to probe the nature of jet magnetic fields in gamma-ray bursts (GRBs). Currently, some GRBs have been detected to be polarized in their early optical afterglows. However, the measured polarization degrees (PDs) of these GRBs are much lower than those predicted by theoretical models. In this work, we investigate the depolarization induced by jet precession in combination with the measured PDs of the GRB early optical afterglows in the reverse shock (RS) dominated phase ($\sim 10^2-10^3 \,{\rm s}$). We calculate the PDs of RS emission with and without jet precession in both magnetic field configurations, i.e., aligned and toroidal magnetic fields, and meanwhile explore the effect of different parameters on the PDs. We find that the PDs are slightly affected by the configurations of the ordered magnetic fields and are positively related to the precession period. Moreover, the PDs are sensitive to the observed angle and the measured low PDs favor a small one. Thus, as one of the plausible origins of the structured jets, jet precession could be considered as an alternative mechanism for the low PDs observed in GRB early optical afterglows.
\end{abstract}


\section{Introduction}

Gamma-ray bursts (GRBs) are the strongest energetic explosions in the Universe and believed to be powered by ultrarelativistic jets \citep[see a review by][]{zhang2018book}. Although GRBs have been studied for several decades, the magnetic field of their jets is still elusive to date. The polarization signal of GRB emission, including both prompt gamma-ray emission and subsequent multiwavelength afterglows, is thought to be a unique tool to probe the magnetic field properties of the jets \citep[for a review, see][]{Gill2021}. For the prompt emission, significant high polarization degrees (PDs) were measured in some GRBs, and thus large-scale ordered magnetic fields from the central engines were expected in jets \citep[e.g.,][]{McGlynn2007,Yonetoku2012}. In addition, a high PD (up to $\sim 60 \, \% $) has been predicted in early afterglows due to the expected ordered magnetic fields \citep[e.g.,][]{Granot2003b}.

As detection techniques have improved over the years, polarization signals from the early optical afterglows of some GRBs have been measured
\citep[e.g.,][]{Mundell2007,Mundell2013,Steele2017,Jordana2021,Shrestha2022,Arimoto2024}. Most of them have a significantly low PD \citep[e.g.,][]{King2014,Buckley2021,Mandarakas2023,Agui2024}.
A small fraction of them shows a relatively high PD, such as GRBs 090102 \citep[$10 \pm 1 \,\%$,][]{Steele2009}, 091208B \citep[$10.4 \pm 2.5 \,\%$,][]{Uehara2012}, 110205A \citep[$13^{+13}_{-9} \,\%$,][]{Steele2017}, and 120308A \citep[$28\pm4\,\%$ to $16^{+5}_{-4}\,\%$,][]{Mundell2013}. However, these measured PDs are still much lower than the theoretical predictions. This means that there are some factors that break down the high polarization arisen from the large-scale ordered magnetic fields, i.e., depolarization effects.

Various explanations have been proposed for the low PDs measured in the early optical afterglows, and the low PDs can generally be classified into external and intrinsic origins. For the external origin, the low PDs are induced by the dust from the Galactic interstellar medium \citep[e.g.,][]{King2014} or from GRB host galaxies \citep[e.g.,][]{Jordana2020}; and for the intrinsic origin, they mainly come from the forward and reverse shocks (FSs and RSs) generated by the collision between the jets and the circumburst medium \citep[e.g.,][]{Gruzinov1999}. Wherein, in the FS frame, the low PDs are caused by the magnetic field generated directly by the FSs \citep[e.g.,][]{Jordana2021, Kuwata2023} or the magnetic field in the circumburst medium compressed by the FSs \citep[e.g.,][]{Teboul2021}. Meanwhile, in the RS frame, if the large-scale ordered magnetic fields from the central engines are not considered to exist in GRB jets, there will be not high polarization in RS emission and the low PDs may be mainly due to the magnetic field originated from the RSs. If considered, the low PDs will be stemmed from the large-scale ordered magnetic field adding one or more depolarization factors, which are usually used to interpret the observed PDs larger than $10\,\%$ \citep[e.g.,][]{Steele2009,Mundell2013}. Here, these depolarization factors can be a high magnetization in the jets \citep[e.g.,][]{Zhang2005,Giannios2008}, an adding tangled magnetic field \citep[e.g.,][]{Lan2019,Tuo2024}, or the direction of the ordered magnetic field partly parallel to the shock normal \citep{Stringer2020}. Besides the magnetic field factors, the changes in jet geometry also affect the PDs \citep[e.g.,][]{Rossi2004,Gill2018}. \cite{Huang2022} proposed that the precession of the jets can significantly weaken the high PD of RS emission with an ordered magnetic field.

In this paper, we present the depolarization caused by the jet precession in combination with the observed data to probe the possible origin of the measured low PDs in the early optical afterglows. The paper is structured as follows. Sections 2 and 3 exhibit the jet precession model and the observed polarization data of the early optical afterglows, respectively. The main results are made in Section 4, and the conclusions and discussion are given in Section 5.

\begin{figure}
\centering
\includegraphics[height=0.8\linewidth, width=0.9\linewidth]{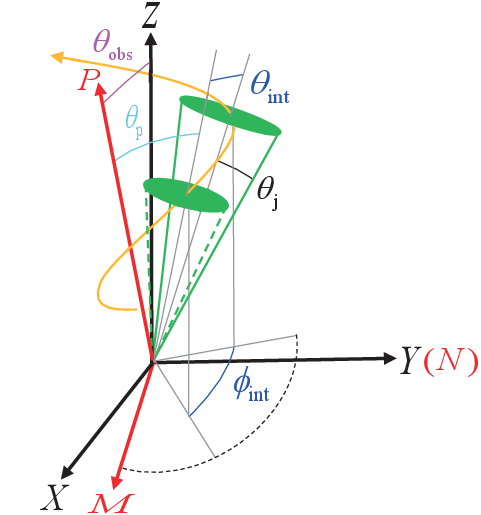}
\caption{Schematic diagram of jet precession.}
\label{Figure1}
\end{figure}

\begin{table*}[!t]
    \footnotesize
    \centering
    \caption{The sample of polarization observations in GRB early optical afterglows.}
    \renewcommand\arraystretch{1.2}
    \setlength{\tabcolsep}{4.0mm}
    \begin{threeparttable}
    \begin{tabular}{cccccc}
    \toprule[0.8pt]
    $\rm GRB$         & $ z$           & $t\,(\rm s)$                  & $\Pi \,(\%)$\tnote{*}  & $E_{\rm \gamma,iso}\, (\rm erg)$\tnote{\dag}       & $\Gamma_{\rm ej}$\\
    \midrule[0.6pt]
        060418        & 1.49                & 203                      & $\textless 8$ [1]       & $10^{53}$ [a]                                     & 324\\
        090102        & 1.547               & $160-220$                & $10\pm 1$ [2]           & $1.4 \times 10^{53}$ [b]                          & 352\\
        091208B       & 1.063               & $149-706$                & $10.4\pm2.5$ [3]        & $2 \times 10^{52}$ [c]                            & 216\\
        100805A       & 1.3                 & $140-320$                & $\textless 14$ [4]      & $7.7 \times 10^{51}$ [f]                          & 170\\
        101112A       & $ \lesssim 3.5$     & $176-355$                & $6^{+3}_{-2}$ [4]       & $5.8 \times 10^{51}$ [f]                          & 158\\
        110205A       & 2.22                & $240-840$                & $13^{+13}_{-9}$ [4]     & $5.6 \times 10^{53}$ [c]                          & 498\\
        110726A       & $1.036< z < 2.7$    & $191-783$                & $\textless 14$ [4]      & $1.2 \times 10^{51}$ [f]                          & 107\\
        120119A       & 1.728               & $194-793$                & $\textless 8$ [4]       & $3.4 \times 10^{53}$ [c]                          & 439\\
        120308A       & $\sim 2.2$          & $240-323$                & $28\pm 4$ [5]           & $2.5 \times 10^{51}$ [f]                          & 146\\
        120311A       & $\lesssim 3$        & $181-779$                & $\textless 13$ [4]      & $1.3 \times 10^{51}$ [f]                          & 109\\
        120326A       & 1.798               & $210-872$                & $\textless 14$ [4]      & $3.6 \times 10^{52}$ [c]                          & 251\\
        140430A       & 1.6                 & $124-424$                & $\textless 22$ [6]      & $1.3 \times 10^{52}$ [d]                          & 222\\
        141220A       & 1.3195              & $129-204$                & $2.8^{+2.0}_{-1.6}$ [7] & $1.8 \times 10^{52}$ [c]                          & 211\\
        151215A       & 2.59                & $182-782$                & $\textless 5.9$ [8]     & $5.6 \times 10^{51}$ [f]                          & 157\\
        180325A       & 2.25                & $147-747$                & $\textless18.2$ [8]     & $3.5 \times 10^{53}$ [f]                          & 443\\
        190114C       & 0.4245              & $203-803$                & $2.9\pm 0.8$ [8]        & $3 \times 10^{53}$  [e]                           & 426\\
    \bottomrule[0.6pt]
    \label{table1}
\end{tabular}
    \quad \textbf{Note.}
    \begin{tablenotes}
    \footnotesize
    \item [*] [1] \cite{Mundell2007}; [2] \cite{Steele2009}; [3] \cite{Uehara2012}; [4] \cite{Steele2017}; [5] \cite{Mundell2013}; [6] \cite{Kopa2015}; [7] \cite{Jordana2021}; [8] \cite{Shrestha2022}.
    \item [\dag] [a] \cite{Butler2007}; [b] \cite{Butler2010}; [c] \cite{Ghirlanda2018}; [d] \cite{Kopa2015}; [e] \cite{MAGIC2019}; [f] $E_{\rm \gamma,iso}$ of other GRBs are calculated by ourselves.
    \end{tablenotes}
\end{threeparttable}
\end{table*}

\section{Jet precession}

In our model, the precession process is represented by a series of discrete sub-jets distributed along the precession path, as shown in Figure \ref{Figure1}. The number of sub-jets per precession period is determined by the angle $\theta_{\rm int}$ between the axes of two adjacent sub-jets and the precession angle $\theta_{\rm p}$, which is the angle between the jet axis and the precession axis. Here, $\theta_{\rm int}$ and $\theta_{\rm p}$ is defined in a coordinate system ($M$,$N$,$P$), where the precession axis being the $P$-axis. Assuming the precession angle $\theta_{\rm p}$ remains constant, the angular separation $\phi_{\rm int}$ between these axes in the azimuthal direction is given by the formula, i.e., $\cos\phi_{\rm int} = (\cos\theta_{\rm int} - \cos^2\theta_{\rm p}) / \sin^2\theta_{\rm p}$. The number of sub-jets in a precession period is then $360^\circ / \phi_{\rm int}$. Additionally, the duration of each sub-jet activity is related to the precession period $T$. The sub-jets in the first period interact with the circumburst medium, generating FSs that sweep up the ISM and RSs that propagate into the sub-jets. As the shocked circumburst medium accumulates, the speed of the resulting blast waves, consisting of the shocked sub-jets and the shocked circumburst medium, decreases. Consequently, sub-jets from subsequent periods catch up with these blast waves, refreshing the RSs and providing additional energy injection.

In this framework, the dynamical model proposed by \cite{Ai2021} is used to first describe the dynamical evolution of the blast waves. The model is developed for the case of a blast wave driven by an ejecta with an arbitrary magnetisation parameter. The dynamics of the blast wave can be described by the ideal magnetohydrodynamic equations. There are four regions separated by a pair of shocks with a contact discontinuity, including the unshocked circumburst medium, the shocked circumburst medium, the shocked ejecta, and the unshocked ejecta. The unshocked circumburst medium is assumed to be non-magnetized and the magnetic field lines in the unshocked ejecta are considered to be parallel to the shock plane. In addition, a constant velocity in the blast wave is assumed. In this model, seven parameters are involved, including the luminosity of the central engine $L_{\rm ej}$, the initial Lorentz factors of the unshocked ejecta $\Gamma_{\rm ej}$, the magnetization parameter of the unshocked ejecta $\sigma_{\rm ej}$, the number density of the unshocked circumburst medium $n$ (interstellar medium considered here, ISM), and the duration of the ejecta $t_{\rm end}$. On the other hand, when the RSs cross the last sub-jets, the shocked ISM and the shocked sub-jets are considered to evolve independently. At this stage, the dynamical evolution of the shocked sub-jets follows the Blandford-McKee self-similar solution, while the evolution of the shocked ISM is described using the model applied in \cite{Chen2021}. The detailed process of dynamical evolution of the blast waves under jet precession can be found in \citet{Huang2022}.

Considering that the radiation mechanism of GRB afterglows is synchrotron emission, the polarization of the synchrotron emission for early optical afterglows is calculated. The total polarization effect results from the sum of the polarization contributions from each blast wave, projected onto the plane of the line of sight. In fact, it is also the superposition of the Stokes parameters in a global coordinate system ($X$,$Y$,$Z$) with the $Z$-axis aligned with the line of sight. But before that, the Stokes parameters from each blast wave calculated in the moving coordinate system should be transferred to the global coordinate system. Due to the symmetry, the polarization from the blast waves will cancel each other out to some extent, and thus the total polarization effect is less significant than that from a single blast wave. The detailed procedure for calculating the polarization of early optical afterglows with jet precession can be found in \citet{Huang2022}.

In this work, we just investigate the depolarization effect of jet precession on high polarization that is due to the large-scale ordered magnetic fields advected from the central engines (i.e., the RS origin), since it is generally believed not high polarization in the FS emission, which is expected to be no or weakly polarized owing to the nearly random direction of magnetic fields in the FS region. For the RSs, the microphysical parameter is the fraction of the internal energy density of RSs assigned to electrons $\varepsilon_{\rm e,r}$. In addition, two configurations are assumed for the large-scale ordered magnetic fields with a coherence scale as large as the jet aperture originating from the central engine, namely aligned magnetic fields (AMF) and toroidal magnetic fields (TMF).

\begin{figure*}[!t]
\centering
\includegraphics[width=0.47\linewidth]{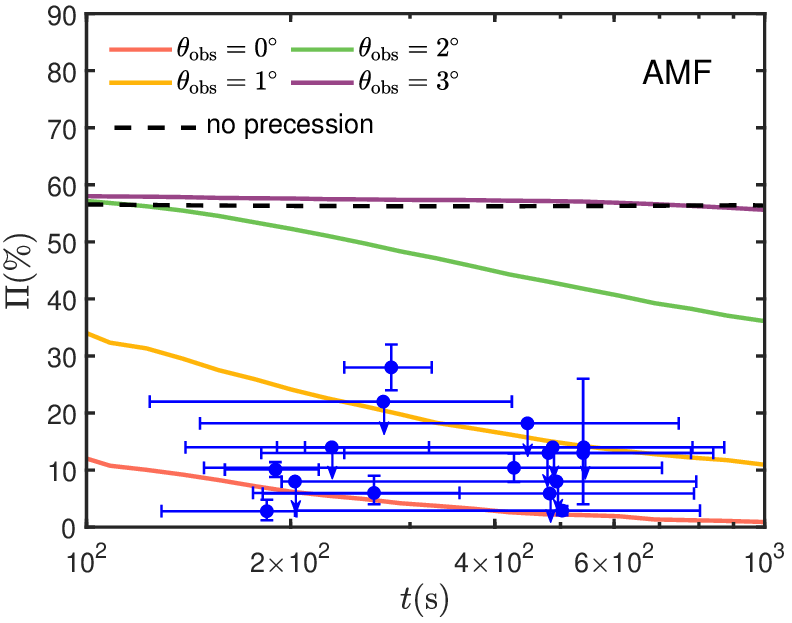}
\includegraphics[width=0.47\linewidth]{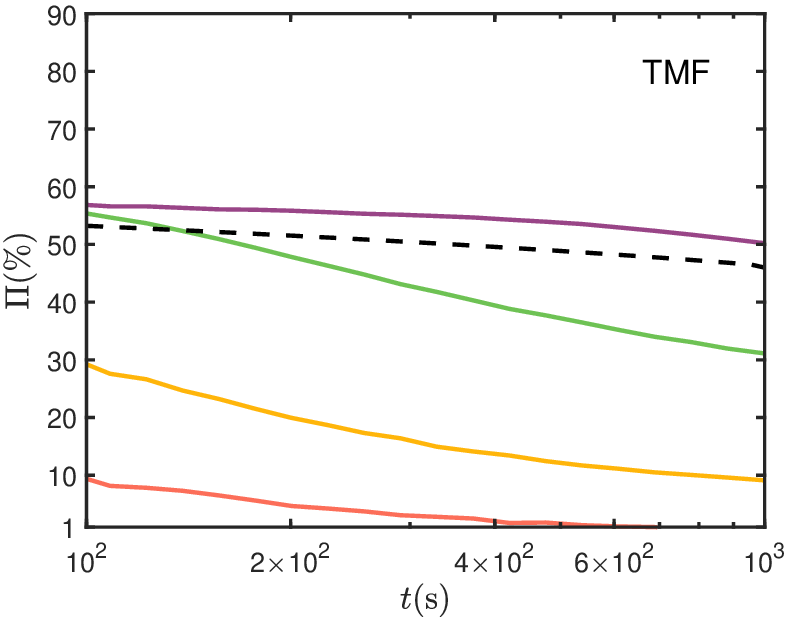}
\caption{PDs evolution of GRB early optical afterglows with observed time. The left and right panels show the results for the AMF and TMF, respectively. The red, yellow, green, and purple lines denote the results of jet precession with $\theta_{\rm obs} = 0^{\circ}$, $1^{\circ}$, $2^{\circ}$, and $3^{\circ}$, respectively, and the black dashed lines indicate the results without jet precession at the viewing angle of $1^{\circ}$. The observed data listed in Table 1 are plotted in the AMF case.}
\label{Figure2}
\end{figure*}

\begin{figure*}[!t]
\centering
\includegraphics[width=1.0\linewidth]{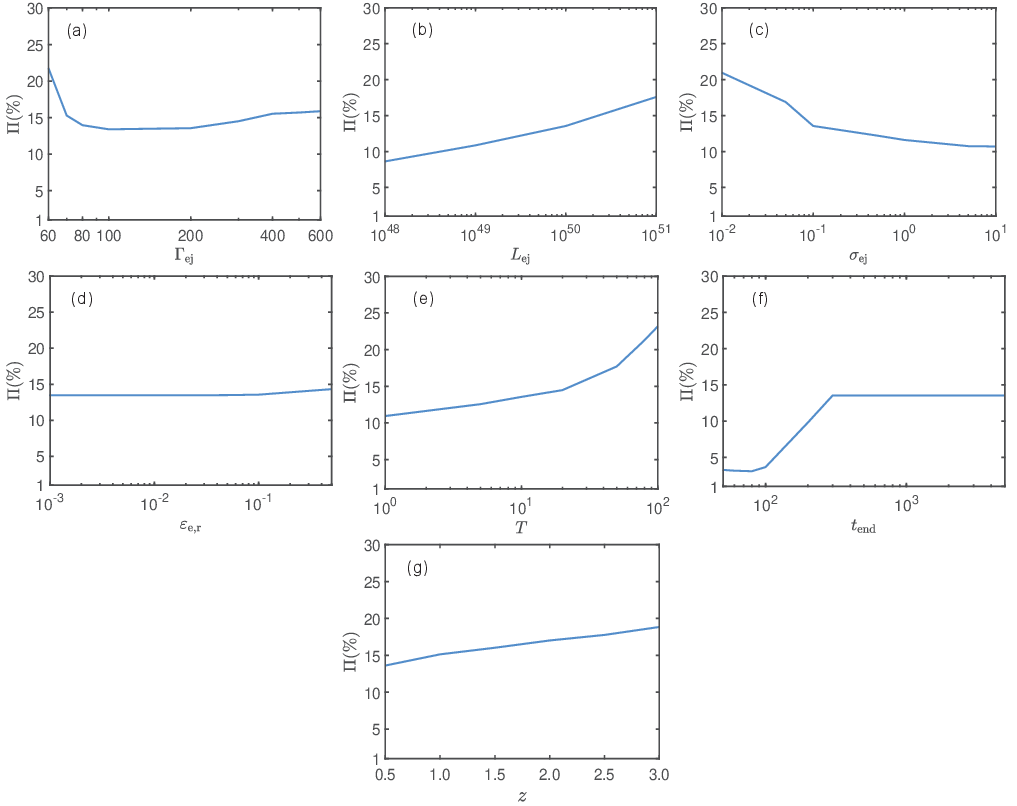}
\caption{PDs of GRB early optical afterglows evolving with different parameters at $\theta_{\rm obs}=1^{\circ}$ and $t=500 \,{\rm s}$, including the initial Lorentz factor $\Gamma_{\rm ej}$, the jet luminosity $L_{\rm ej}$, the magnetization parameter $\sigma_{\rm ej}$, the fraction of the internal energy density in the RS downstream shared by electrons $\varepsilon_{\rm e,r}$, the precession period of the jets $T$, the duration of the jets $t_{\rm end}$, and the redshift $z$.}
\label{Figure3}
\end{figure*}

\section{Measured PDs of early optical afterglows}\label{s3}

Based on lightcurve observations \citep[e.g.,][]{Japelj2014,Yi2020}, the RSs are considered to be dominated in the early phase of GRB afterglows ($<10^3 \, {\rm s}$), and its radiation appears mainly in optical bands. Considering that the measured PDs of early optical ($<10^2 \,{\rm s}$) could be the prompt emission origin, we collect the observed PDs of early optical in the time interval of $10^2-10^3\,{\rm s}$ as the sample studied in this work. 16 GRBs are collected, of which 7 GRBs (GRBs 090102, 091208B, 101112A, 110205A, 120308A, 141220A, and 190114C) have a definite PD and the others only have a PD upper limit. These GRBs' names, redshifts, PDs, corresponding observation times, isotopic gamma-ray energies, and initial Lorentz factors are listed in Table \ref{table1}. GRBs 090102, 091208B, 110205A, and 120308A have a PD equal to or greater than $10 \,\%$. For GRB 090102, a $10\pm1 \, \%$ PD was measured in time of $160-220 \,{\rm s}$, then \citet{Steele2009} confirmed the presence of a large-scale ordered magnetic field in its jets. For GRB 091208B, a PD of $10.4\pm2.5 \, \%$ was measured in the early optical afterglow phase, but it was considered to be the FS origin, because of the corresponding optical light curve with a power-law decay index, $-0.75 \pm 0.02$, in agreement with the prediction of the FS model \citep{Uehara2012}. However, the mechanism of the large-scale ordered magnetic field produced in the FS region is still unclear, and thus the origin of this PD remains under debate, so that the RS origin is not excluded. Similarly, a $13^{+13}_{-9}\,\%$ PD has been measured in the early optical afterglows of GRB 110205A, but the PD has a significantly larger error than those in GRBs 090102 and 091208B. In addition, GRB 120308A shows an optical PD that evolves from $28\pm4 \, \%$ to $16^{+5}_{-4} \, \%$, which is interpreted as the increasing contribution of the FS emission to the total emission with time \citep{Mundell2013}. Here, we select only the largest value $28\pm4 \, \%$ corresponding to the peak of the RS light curve. Although the high PD was measured in the early optical afterglow of GRB 120308A, suggesting the existence of magnetic fields with a sufficiently large coherence scale in its jets, similar events of such high polarisation are quite rare. This implies the mechanisms that can modify the geometry of the magnetic field or the jet structure may play an efficient role in eliminating the high PDs for the early optical afterglows dominated by the RSs.

For obtaining the distribution of the measured PDs in the parameter space of the initial Lorentz factors, we used an empirical correlation proposed by \cite{Liang2010} to roughly estimate the initial Lorentz factors of these GRBs, i.e., $\Gamma_{\rm ej} \simeq 182(E_{\rm \gamma,iso}/10^{52} \, {\rm erg})^{0.25}$, with $E_{\rm \gamma,iso}$ representing the isotropic prompt gamma-ray energy of GRBs. Here, $E_{\rm \gamma,iso}$ of part of these GRBs are obtained from previous literatures and the others are calculated by ourselves. For GRBs marked by [f] in Table 1, GRB 101112A was detected by Fermi and others were observed by Swift, whose energy band is extended to $10^4 \, {\rm keV}$.

\section{Result}

We calculate the PDs of the RS emission in the optical band with jet precession. We consider a top-hat jet with a half-opening angle $\theta_{\rm j}$ without sideways expansion and choose a set of fiducial values for the parameters, including $L_{\rm ej} = 10^{50} \, {\rm erg\,s^{-1}}$, $\Gamma_{\rm ej} = 200$, $\sigma_{\rm ej} = 0.1$, $n=1\,{\rm cm^{-3}}$, $\theta_{\rm j} = 5^{\circ}$, $T = 10 \,{\rm s}$, $t_{\rm end} = 10^3 \,{\rm s}$, and $\varepsilon_{\rm e,r}=0.1$. Meanwhile, we set $\theta_{\rm p} = \theta_{\rm j}$ and $\theta_{\rm obs}= 0^{\circ}$, $1^{\circ}$, $2^{\circ}$, and $3^{\circ}$, respectively. $T$ and $t_{\rm end}$ are in the frame of the burst source. The observed angle $\theta_{\rm obs}$ is the angle between the line of sight and the precession axis as shown in Figure 1. In addition, considering that the precession path should be completely filled by the discrete sub-jets and that the overlap between adjacent sub-jets should be as small as possible, we set $\theta_{\rm int}= 7^{\circ}$, thus obtaining 4 sub-jets in one precession period. For a clear comparison, we also calculate the PDs of the RS emission in the optical band without jet precession.

\begin{figure*}
\centering
\includegraphics[width=0.47\linewidth]{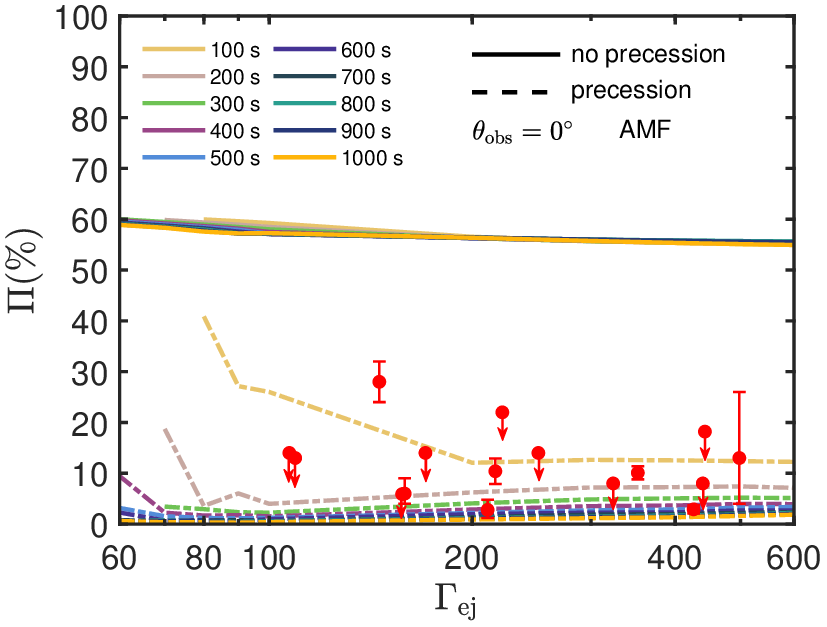}
\includegraphics[width=0.47\linewidth]{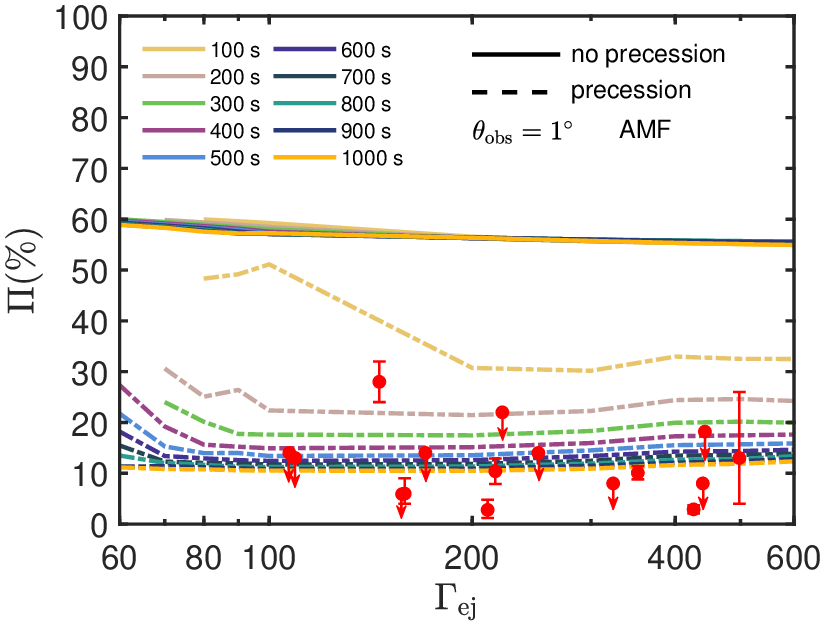}
\caption{PDs versus Lorentz factors in the GRB early optical afterglows. The left and right panels show the results with $\theta_{\rm obs} = 0^{\circ}$ and $1^{\circ}$, respectively. The estimated Lorentz factors are listed in Table \ref{table1}. Different color lines represent the results at different times from $10^2 \,{\rm s}$ to $10^3 \,{\rm s}$, where dashed and solid lines denote the results with and without jet precession, respectively.}
\label{Figure4}
\end{figure*}

\subsection{Theoretical PDs under jet precession}

Figure \ref{Figure2} displays the evolution of the PDs over observed time during the RS dominated phase. The left and right panels depict the results for the AMF and TMF, respectively. The red, yellow, green, and purple solid lines correspond to the results of $\theta_{\rm obs} = 0^{\circ}$, $1^{\circ}$, $2^{\circ}$, and $3^{\circ}$, respectively, while the black dashed lines represent the scenario without jet precession for a viewing angle of $1^{\circ}$. Note that the viewing angle without jet precession is the angle between the line of sight and the jet axis.

The figure demonstrates that the PDs are highly sensitive to $\theta_{\rm obs}$ in the jet precession scenario. As $\theta_{\rm obs}$ decreases, PDs also decrease significantly. Specifically, at $\theta_{\rm obs} = 3^\circ$, the PDs are approximately $60\,\%$ close to the scenario without jet precession, whereas at $\theta_{\rm obs} = 0^\circ$, the PDs drop to below $10\,\%$. Of course, the PDs are also related to other parameters of the geometry, dynamics, and radiation mechanism of jets. Additionally, PDs decay noticeably over time. For instance, at $\theta_{\rm obs} = 1^\circ$, PDs decrease from $30\,\%$ to $10\,\%$ in the time interval of $10^2-10^3 \,{\rm s}$. This decay is not found in the scenario without jet precession. Furthermore, there is no significant difference between the AMF and TMF results, indicating that the type of ordered magnetic field configurations not substantially affect PD evolution in the presence of jet precession. Overall, the depolarization effect due to the jet precession significantly impacts the high PDs originating from the RSs with large-scale ordered magnetic fields, particularly when $\theta_{\rm obs}$ is less than $2^{\circ}$ due to the symmetry in the ejecta structure.

We also investigated the effect of various parameters on the PDs. These parameters include the initial Lorentz factor $\Gamma_{\rm ej}$, the jet luminosity $L_{\rm ej}$, the magnetization parameter $\sigma_{\rm ej}$, the fraction of the internal energy density in the downstream RS shared by electrons $\varepsilon_{\rm e,r}$, the jet precession period $T$, the jet duration $t_{\rm end}$, and the redshift $z$. The results are displayed in Figure \ref{Figure3}. In these figures, we assume the AMF, $\theta_{\rm obs} = 1^\circ$, and $t = 500 \, {\rm s}$. Each panel shows the PD evolution with respect to one parameter, while the other parameters are held constant at their fiducial values as displayed in the beginning of Section 4.

For the initial Lorentz factor $\Gamma_{\rm ej}$, decreasing from 600 to 80, the PD shows minimal variation, but a significant increase in PD occurs when $\Gamma_{\rm ej}$ drops below 80 as shown in Figure 3(a). Here, the PD is obtained by summing the contributions from the RS of each blast wave system generated by interactions between the sub-jets in the first periodic cycle and the ISM at the observer time $t = 500 \, {\rm s}$. Due to the location symmetry of partial radiation emitters among these blast wave systems relative to the line of sight, high polarization from each blast wave system is partially counteracted each other and the extent of which depends mainly on the direction of the line of sight (i.e., the observer angles) and the dynamic states of these blast wave systems. Given the observer angles, for relatively large $\Gamma_{\rm ej}$, the RSs of the systems form immediately after the interactions and their dynamic states have been tended to be similar at $t = 500 \, {\rm s}$. In this case, variations in $\Gamma_{\rm ej}$ have minimal impact on the PD. However, when $\Gamma_{\rm ej}$ is sufficiently small, the RSs do not form immediately after the interactions and the delay time increases as $\Gamma_{\rm ej}$ decreases. In this case, the dynamic state of the RS of each blast wave system differ more noticeably at $t = 500 \, {\rm s}$, leading to more pronounced anisotropy among them regarding the line of sight, and thus the counteraction effect becomes weaken.

In Figure 3(b), the PD increases gradually with rising $L_{\rm ej}$, while in Figure 3(c), the PD decreases as $\sigma_{\rm ej}$ increases. For the former, this is because the distances of the blast wave systems from the centre at $t = 500 \, {\rm s}$ become large with increasing $L_{\rm ej}$ and thus the differences among the distances of the systems become larger, resulting in an improvement in the anisotropy level. For the latter, the dynamic model incorporates a magnetic field component, meaning that variations in $\sigma_{\rm ej}$ can affect the dynamic evolutions of the blast wave systems. The effect of $\sigma_{\rm ej}$ on the polarization behaves inversely compared to $L_{\rm ej}$. Figure 3(d) shows that the PD remains constant with increasing $\varepsilon_{\rm e,r}$. This is because $\varepsilon_{\rm e,r}$, the fraction of the internal energy density of the RSs assigned to the shock-accelerated electrons, is considered not to influence the dynamic evolutions of the blast wave systems in the model. In Figure 3(e), the PD increases with a longer precession period $T$. A longer precession period results in greater differences in distance between adjacent blast wave systems, improving the degree of the anisotropy and resulting in a higher PD after counteraction. In Figure 3(f), the PD increases rapidly before the duration $t_{\rm end}$ reaches around $250\, {\rm s}$, beyond which the PD stabilizes. The longer $t_{\rm end}$, the longer the duration of the energy injection into the blast wave systems, and thus the greater the differences among them, leading to an increase in the PD. But here the increase of $t_{\rm end}$ has no effect on the PD anymore when $t_{\rm end}(1+z)>500\,{\rm s}$ because the PD is adopted at the observer time $t=500\,{\rm s}$. Finally, Figure 3(g) shows the PD evolving with the redshift, revealing slight PD variation with the increasing redshift. This is due to the magnification effect of redshift on observer time and not an intrinsic change in polarization level caused by the dynamic evolutions like those mentioned above. Given the observer time $t=500\,{\rm s}$ with the redshift considered, the observer time without the redshift considered $t/(1+z)$ becomes small for the increasing redshift, and the corresponding PD becomes large, since PDs increase with decreasing observer time in the scenario of jet precession. Here, the redshift $z=1$ is adopted in the theoretical calculations. In this paper, we assume that the jet structure is uniform, i.e., a top-hat jet. For a structured jet, the polarisation level could be further reduced because in the structured jet scenario, the energy density in the jet core will be significantly larger than that at the jet edge and the PDs are proportional to the jet lunimosity as shown in Figure 3(b). In summary, variations in these parameters can affect the PD to some extent, with fluctuations ranging approximately from $\sim 10\,\%$ to $\sim 20\,\%$. This range is comparable to the PD variation over time with $\theta_{\rm obs} = 1^\circ$ (see Figure \ref{Figure2}), indicating that the observer angle, $\theta_{\rm obs}$, has a more substantial impact on PDs than other parameters.

Jet precession can be considered as a natural mechanism to produce the structured jets, and its depolarization effect is significant for a relatively small $\theta_{\rm obs}$ with the large-scale ordered magnetic fields.

\subsection{Implications from the measured PDs}

The left panel of Figure \ref{Figure2} also illustrates the measured PDs discussed earlier in Section \ref{s3}, represented by blue circles. The downward arrows indicate upper limits on the PDs. This comparison aims to explore the potential origin of the observed PDs through jet precession, based on a sample analysis. It is evident that the measured PDs generally support a small angle of $\theta_{\rm obs}$ around $1^{\circ}$, indicating that jet precession might be a plausible explanation for the low PDs observed in early optical afterglows. However, we cannot conclusively attribute the low PDs of any specific a-GRB to jet precession due to the lack of direct observational evidence of jet precession.

Figure \ref{Figure4} presents the measured PDs within the parameter space defined by the initial Lorentz factors $\Gamma_{\rm ej}$, alongside PD curves predicted theoretically. The red circles denote the measured PDs, with downward arrows indicating upper limits. The solid and dashed lines represent theoretical PDs without and with the jet precession effect, respectively, using different colors to depict results at ten time points ranging from $10^2$ to $10^3$ s. The left and right panels exhibit the theoretical PDs calculated for $\theta_{\rm obs} = 0^\circ$ and $1^\circ$, respectively. The analysis assumes the AMF and the fiducial values for all parameters except for the initial Lorentz factors, which are detailed in Table \ref{table1}.

From Figure \ref{Figure4}, it is apparent that there is no significant trend in the measured PDs when they relate to the initial Lorentz factors. Likewise, the theoretical PDs, irrespective of the jet precession effect, generally remain constant except at $t = 10^2$ s in the jet precession scenario. Notably, the theoretical PDs without jet precession are significantly higher than the measured PDs. However, the theoretical PDs accounting for jet precession encompass most of the measured PDs, particularly when $\theta_{\rm obs}$ is close to zero.

\section{Conclusions and discussion}

Observations show that the measured PDs in the early optical afterglows of GRBs are significantly lower than the high PDs predicted by early models. The cause of these low PDs remains unclear. Given that jet precession is expected to exist in GRBs, we explore the possibility of jet precession as a potential explanation for the observed low PDs.

Structured jets are crucial for studying the afterglow light curves of short GRBs associated with gravitational waves (GWs), as exemplified by GRB 170817A \citep{Abbott2017}. Various jet structures have been proposed, including the power-law structures \citep[e.g.,][]{Granot2003a}, the Gaussian structures \citep[e.g.,][]{Zhang2002}, the quasi-spherical structures \citep[e.g.,][]{Gill2018}, and the asymmetric multi-zone structures \citep[e.g.,][]{Li2023}. However, the physical origins of these structures remain unclear; they may arise from a top-hat jet with an additional cocoon or wind component. We consider that these structures could result from the precession of a top-hat jet, as the precessional motion of the jet resembles the reshaping process of the jet \citep{Huang2019}. Currently, these potential origins cannot be distinguished solely by light curves. Observing the polarization in early afterglows may provide a way to differentiate them, as early polarization signals can reveal the magnetic field properties of the ejecta. The jets powered by the central engine, the cocoon typically formed as the jets penetrate the merger's ejecta \citep[e.g.,][]{Gottlieb2018}, and the wind potentially originating from a newborn magnetar \citep[e.g.,][]{Dai1998,Song2023} or a black hole hyperaccretion system \citep[e.g.,][]{Liu2017,Liu2018}, should exhibit distinct magnetic field characteristics. For short GRBs associated with GWs, viewing angles can be determined through GW observations, and jet structures can be constrained by the light curves of multiband afterglows. Consequently, the magnetic field properties of the jets can be inferred from polarization signals, providing an insight into the origin of jet structures, if high-quality polarization observations are available.

In this paper, we focus solely on RS emission. However, FSs in fact occur simultaneously with RSs during the RS dominated phase. If we consider an FS emission contribution to the total emission during this phase, the high polarization typically associated with RSs with large-scale ordered magnetic fields will further decrease. The extent of this decrease depends on the ratio of FS and RS emissions. Additionally, while we discuss polarization measurements primarily in the optical band, polarization can also be measured in other bands. For example, GRB 190114C exhibited polarization in the radio/millimeter band, varying from $0.87\,\%$ to $0.6\,\%$ \citep{Laskar2019}. Furthermore, an upper limit of $13.8\,\%$ in X-ray polarization was posed for GRB 221009A by the Imaging X-ray Polarimetry Explorer \citep[IXPE,][]{Negro2023}. Considering the relatively high PDs in prompt emission of GRBs, the evolution of jet precession should be investigated to outline the evolution of PDs in whole process of GRBs.

\acknowledgments
This work was supported by the National Key R\&D Program of China under grants 2023YFA1607902, the National Natural Science Foundation of China under grants 12173031 and 12221003, and the Fundamental Research Funds for the Central Universities (No. 20720240152).

\end{document}